\documentclass[aps,prd,twocolumn,groupedaddress]{revtex4}
\usepackage[dvips]{graphicx}

\def\sci#1#2{#1\times 10^{#2}}
\newcommand{\degree}{^\circ}
\newcommand{\sign}{\textrm{sign}}
\newcommand{\fcep}{f_{\parallel,ce}}
\newcommand{\E}{\mathbf{E}}
\newcommand{\En}{\mathcal{E}}
\newcommand{\Enu}{\En_\nu}
\newcommand{\Elpm}{\En_{\rm LPM}}

\newcommand{\tlpm}{t_{\rm LPM}}
\newcommand{\smax}{s_{\rm up}}
\newcommand{\Rsat}{R_{\rm sat}}
\newcommand{\hsat}{h_{\rm sat}}
\newcommand{\Rearth}{R_\oplus}
\newcommand{\fluxfraction}{F}
\newcommand{\nadir}{b}
\newcommand{\emang}{\theta}
\newcommand{\polang}{p}
\newcommand{\dipang}{\alpha_{\rm dip}}
\newcommand{\ths}{\theta_s}
\newcommand{\phs}{\phi_s}
\newcommand{\pha}{\phi_a}
\newcommand{\azang}{a}
\newcommand{\antang}{\alpha}

\begin{document}
\title{FORTE satellite constraints \\
on ultra-high energy cosmic particle fluxes}

\author{Nikolai G. Lehtinen}
\author{Peter W. Gorham}
\altaffiliation[Joint appointment with ]{Jet Propulsion Laboratory,
  4800 Oak Grove Dr., Pasadena, CA 91109}
\affiliation{University of Hawaii at Manoa, Dept. of Physics and
Astronomy, Honolulu, HI 96822, U.S.A.}
\author{Abram R. Jacobson}
\author{Robert A. Roussel-Dupr\'e}
\affiliation{Los Alamos National Laboratory, Los Alamos, NM 87545,
  U.S.A.} 
\date{\today}

\begin{abstract}
The FORTE (Fast On-orbit Recording of Transient Events) satellite
records bursts of electromagnetic waves arising from near the Earth's
surface in the radio frequency (RF) range of 30 to 300~MHz with a dual
polarization antenna. We investigate the possible RF signature of
ultra-high energy cosmic-ray particles in the form of coherent
Cherenkov radiation from cascades in ice. We calculate the sensitivity
of the FORTE satellite to ultra-high energy neutrino (UHE $\nu$)
fluxes at different energies beyond the Greisen-Zatsepin-Kuzmin (GZK)
cutoff.  Some constraints on supersymmetry model parameters are also
estimated due to the limits that FORTE sets on the UHE neutralino
flux.  The FORTE database consists of over 4~million recorded events
to date, including in principle some events associated with UHE $\nu$.
We search for candidate FORTE events in the period from September 1997
to December 1999. The candidate production mechanism is via coherent
VHF radiation from a UHE $\nu$ shower in the Greenland ice sheet.  We
demonstrate a high efficiency for selection against lightning and
anthropogenic backgrounds. A single candidate out of several thousand
raw triggers survives all cuts, and we set limits on the corresponding
particle fluxes assuming this event represents our background level.
\end{abstract}

\maketitle

\section{Introduction}
\label{sec:introduction}
Detection and modeling of the highest energy cosmic rays and the
neutrinos which are almost certain to accompany them represents one of
the most challenging problems of modern physics. To date a couple of
tens of cosmic ray events, presumably protons, have been observed with
energies in excess of 10$^{20}$~eV.  The origin of this flux
represents a puzzle since above $\sim\!\sci{5}{19}$~eV the cosmic ray
flux is expected to be reduced due to Greisen-Zatsepin-Kuzmin
(GZK)~\cite{Greisen66,Zatsepin66} mechanism.  The primary particles
inevitably generate ultra-high energy neutrinos (UHE $\nu$) in cosmic
beam dumps.  Weakly interacting neutrinos, unlike gamma photons or
protons, can reach us from distant sources and therefore are a
possible invaluable instrument of high-energy astrophysics.

Above $\sim 10^{19}$~eV, charged cosmic rays are no longer magnetically
confined to our galaxy. This implies that particles above this
energy detected at earth are very likely to be produced in extragalactic
astrophysical sources. Furthermore, the existence of particles up to,
and possibly beyond the $\sim 10^{19.5}$~eV endpoint of the allowed
energy spectrum for propagation over cluster-scale distances suggests
that there is no certain cutoff in the source energy spectra. In fact, we
are unlikely to learn about the endpoint in the source energy
spectra via either charged particles or photons: the universe is largely
opaque to all such traditional messengers.

There are no clear physics constraints on source energy spectra until
one reaches Grand-Unified Theory (GUT) scale energies near
$10^{24}$~eV where particle physics will dominate the production
mechanisms.  If we assume that measurements at $\sim$10$^{20}$~eV
represent mainly extragalactic sources, then there are virtually no
bounds on the intervening three or four decades of energy, except what
can be derived indirectly from upper limits at lower energies. For
these reasons, measurements which are sensitive to ultra-high energy
neutrinos are of particular importance in understanding the ultimate
limits of both particle acceleration in astrophysical zevatrons
(accelerators to $\sim\!\textrm{ZeV}=10^{21}$~eV and higher energies)
and top-down decay of exotic forms of cosmic energy.
 Even at energies
approaching the GUT scale, the universe is still largely transparent
to neutrinos. Also, the combination of a slowly increasing
cross-section combined with the large energy deposited per interaction
makes such neutrino events much more detectable than at lower
energies.

The chief problem in UHE $\nu$ detection arises not from the character
of the events, but from their extreme rarity. For neutrinos with
fluxes comparable to the extrapolated ultra-high energy cosmic ray
flux at $10^{21}$~eV, a volumetric aperture of order
10$^6$\,km$^3$\,sr (water equivalent) is necessary to begin to achieve
useful sensitivity. Such large volumes appear to exclude embedded
detectors such as AMANDA~\cite{AMANDA} or IceCube~\cite{IceCube},
which are effective at much lower neutrino energies.
Balloon-based~\cite{ANITA} or space-based~\cite{EUSO,OWL} systems
appear to be the only viable approaches currently being implemented.

The most promising new detection methods appear to be those which
exploit the coherent radio Cherenkov emission from neutrino-induced
electromagnetic cascades, first predicted in the 1960's by
Askaryan~\cite{Askaryan62,Askaryan65}, and confirmed more recently in
a series of accelerator experiments~\cite{AccelAsk1,AccelAsk2}. Above
several PeV ($10^{15}$~eV) of cascade energy, radio emission from the Askaryan
process dominates all other forms of secondary emission from a
shower. At ZeV energies, the coherent radio emission produces pulsed
electric field strengths that are in principle detectable even from
the lunar surface~\cite{GLUE1,GLUE2}.

These predictions, combined the strong experimental support afforded by
 accelerator measurements, are the basis for our efforts to search
existing radio-frequency data from the Fast On-orbit Recorder of
Transient Events (FORTE) satellite for candidate ZeV neutrino events.
Here we report the first results of this search, based on analysis of
several days of satellite lifetime over the Greenland ice sheet.

\section{Detector characteristics}
\label{sec:forte}
The experiment described in this paper is based on detection of
electromagnetic emission generated in the Greenland ice sheet by the FORTE
satellite.

The FORTE satellite~\cite{Jacobson99} was launched on August 29, 1997
into a 70$\degree$ inclination, nearly circular orbit at an altitude
of 800~km (corresponding to a field of view of $\sim\!27^\circ$ arc
distance). The satellite carries two broadband radio-frequency (RF)
log-periodic dipole array antennas (LPA) that are orthogonal to each
other and are mounted on the same boom pointing in the nadir
direction. The antennas are connected to two radio receivers of 22~MHz
bandwidth and center frequency tunable in 20--300~MHz range. Beside RF
receivers, the satellite carries an Optical Lightning System (OLS)
consisting of a charge-coupled device (CCD) imager and a fast
broadband photometer.  Although for this paper we do not report
analysis of optical data, in more detailed studies the optical
instrument data can be used to correlate RF and optical
emissions~\cite{Suszcynsky00}.

The satellite recording system is triggered by a subset of
8~triggering subbands which are spaced at 2.5~MHz separations and are
1~MHz wide. The signal has to rise 14--20~dB above the noise to
trigger. Typically, a trigger in 5 out of 8 subbands is required.  The
triggering level and algorithm can be programmed from the ground
station.  Multiple channels are needed for triggering because of
anthropogenic noise, such as TV and FM radio stations and radars,
which produce emission in narrow bands which can coincide
with a trigger subband.  After the trigger, the RF data is digitized
in a 12-bit Data Acquisition System (DAS) at 50~Msamples/s, and the
typical record length is 0.4~ms.  The FORTE database consists of over
4~million events recorded in the period from September 1997 to
December 1999.

The ice with its RF refraction coefficient of
$n\approx1.8$~\cite{Johari75} and Cherenkov angle of
$\theta_{C}\approx55.8^\circ$ and relatively low electromagnetic wave
losses in the radio frequency range is a good medium for exploiting the
Askaryan effect for shower detection. The biggest contiguous ice
volume on Earth (the Antarctic) is unfortunately not available to
FORTE satellite because of its orbit inclination of $70^\circ$. The
next biggest contiguous ice volume is the Greenland ice sheet.  Its
area is $\sci{1.8}{6}$~km$^2$, and the depth is $\sim$3~km at the
peak. However, the available depth is limited by RF
losses~\cite{Bogorodsky85} to 1~km. Thus the volume of
Greenland ice observed from orbit is $\approx\sci{1.8}{6}$~km$^3$.

\section{Cherenkov radio emission from particle showers}
\label{sec:cherenkov}
We define the electric field pulse spectrum as
$\mathbf{E}(\omega)=2\int_{-\infty}^{+\infty}\mathbf{E}(t) e^{i\omega t}dt$.
An empirical formula for $\mathbf{E}(\omega)$
from an electromagnetic shower in ice was obtained by Zas et al. \cite{Zas92}:
\begin{eqnarray}
R|\E(\omega)|=\sci{1.1}{-7}\frac{\En_{\rm shower}}{1{\rm\
TeV}}\frac{f}{f_0} \frac{1}{1+0.4(f/f_0)^2}\times{}\nonumber\\
{}\times e^{-\frac{(\emang-\theta_c)^2}{2\Delta\theta}}
{\rm\ V\,MHz^{-1}}
\label{eq:zhs}
\end{eqnarray}
where $R$ is the distance to the observation point in ice, $\En_{\rm
shower}$ is the shower energy, assumed to be $\alt$1~PeV, $f$ is the
electromagnetic wave frequency, $f_0=500$~MHz, $\theta_c=55.8^\circ$
is the Cherenkov angle, and $\Delta\theta=2.4^\circ f_0/f$.

The FORTE detector triggers whenever the amplitude of the electric
field after a narrow-band (1 MHz) filter exceeds a set threshold (in
several channels). Since $|\E(\omega)|$ varies slowly enough in this
band, the peak value of $E$ on the filter output equals
$|\E(\omega)|\Delta f$, where $\Delta f$ and $f=\omega/(2\pi)$ are
correspondingly the bandwidth and the central frequency of the filter.

For low frequencies the emission cone is broad
($\Delta\theta\sim$~radian) and the empirical formula (\ref{eq:zhs})
is not very accurate. Instead, we make an analytical estimate for
emission in Appendix~\ref{app:cherenkov} and get
\begin{equation}
R|\E(\omega)|=\sqrt{2\pi}\mu\mu_0 Q L f
 \sin\theta e^{-(kL)^2(\cos\theta-1/n)^2/2}
\label{eq:cher}
\end{equation}
where $k=2\pi nf/c$ and $n=1.8$ is the ice refraction coefficient, and
$\mu=1$.  At $kL\gg1$ this formula matches the empirical formula
(\ref{eq:zhs}) for $L\approx1.5$~m and $Q\approx(\En_{\rm shower}/1{\rm\
TeV})\times\sci{5.5}{-17}$~C, which agrees with the results of shower
simulations in Figures~1,~2 in \cite{Zas92}. Equation (\ref{eq:cher})
also matches the numerical result for the radiation pattern at 10~MHz
(Figures~11,~12 in \cite{Zas92}) better than the empirical formula
(\ref{eq:zhs}) which becomes inaccurate at low frequencies.

Note that we cannot use equations (\ref{eq:zhs}), (\ref{eq:cher}) for
electromagnetic showers started by particles of high energies because
of significant elongation due to 
 Landau-Pomeranchuk-Migdal (LPM)
effect~\cite{Landau44,Landau52,Landau53,Migdal56}.  According to
\cite{Stanev82}, the LPM effect is important for particle energies
$\En>\Elpm$, where $\Elpm=2.4$~PeV. According to
Appendix~\ref{app:lpm}, for electromagnetic showers with starting
energy $\En_0\gg\Elpm$, the electric field at the exact Cherenkov
angle is still approximately given by (\ref{eq:zhs}), while the width
of Cherenkov cone is reduced to
\begin{equation}
\Delta\theta_{\rm LPM}\approx 0.9^\circ \frac{1}{\sqrt{\En_0/1{\rm\ EeV}}}
\frac{f_0}{f}
\end{equation}
where $f_0=500$~MHz.  For this reason, pure electromagnetic showers
(primarily from $\nu_e$ or $\bar{\nu}_e$ charged current interactions) make a
negligible contribution to the FORTE sensitivity because, as we will
see, the energy threshold is $\gg$1~EeV.

After the interaction with a nucleon, the UHE $\nu$ energy goes into
leptonic and hadronic parts. In the case of an electron
(anti-)neutrino $\nu_e$ ($\bar{\nu}_e$), the electron created in the
charged current interaction starts an electromagnetic shower, which is
however very elongated due to LPM effect making it virtually
undetectable.  In the case of a $\nu_\mu$ ($\bar{\nu}_\mu$) or
$\nu_\tau$ ($\bar{\nu}_\tau$) the lepton does not start a shower.  The
muons and tau leptons deposit their energy due to
electromagnetic~\cite{Mitsui92} and photonuclear~\cite{Bugaev03}
interactions. However, the portions in which the energy is deposited
are not enough to be observed by FORTE. Also, the tau lepton may decay
at a long distance from its creation point after its energy is reduced
to $\alt$1~EeV (10$^{18}$~eV). Thus, tau lepton decay is also
unobservable by FORTE.

On the basis of these arguments, we only consider the
neutrino-initiated hadronic shower, and we do not distinguish between
neutrino flavors. Most of the hadronic energy converts in the end into
electromagnetic, thus Cherenkov emission as a result of Askaryan
effect can be used for its detection. Moreover, the LPM effect does
not produce significant shower elongation, as shown by Alvarez-Mu\~niz
and Zas~\cite{Alvarez98}, since most of the $\pi_0$ particles which
decay into photons instead of interacting have their energy reduced
below the LPM level. Thus, the value of $L\approx 1.5$~m used in
(\ref{eq:cher}) does not change appreciably. This is in accordance
with results of Monte Carlo calculations of~\cite{Alvarez98}, showing
that the Cherenkov cone narrows by only $\sim$30\% when hadronic
shower energy ranges from 1~TeV to 10~EeV.

The hadronic shower energy which should be used in (\ref{eq:zhs}) or
(\ref{eq:cher}) is $\En_{\rm shower}=y\Enu$, where $\Enu$ is the
neutrino energy, and $y$ is the fraction of energy going into the
hadronic shower. The theoretical value for UHE $\nu$ is $\langle
y\rangle\approx 0.2$~\cite{Gandhi96}.

Because the long wavelength observed by FORTE ($\sim$6~m in ice) are
far greater than the Moli\`ere (transverse) radius of the showers
involved ($\sim$11.5~cm), the transverse structure of the shower
(expressed by the factor $(1+0.4(f/f_0)^2)^{-1}$ in (\ref{eq:zhs})) is
neglected in our analysis. The same argument applies to transverse
non-uniformities of hadronic showers.

\section{Search for relevant signals in the FORTE database}
\subsection{Geographic location of FORTE events}
\label{ssec:geolocation}
\begin{figure*}
\includegraphics[width=7in]{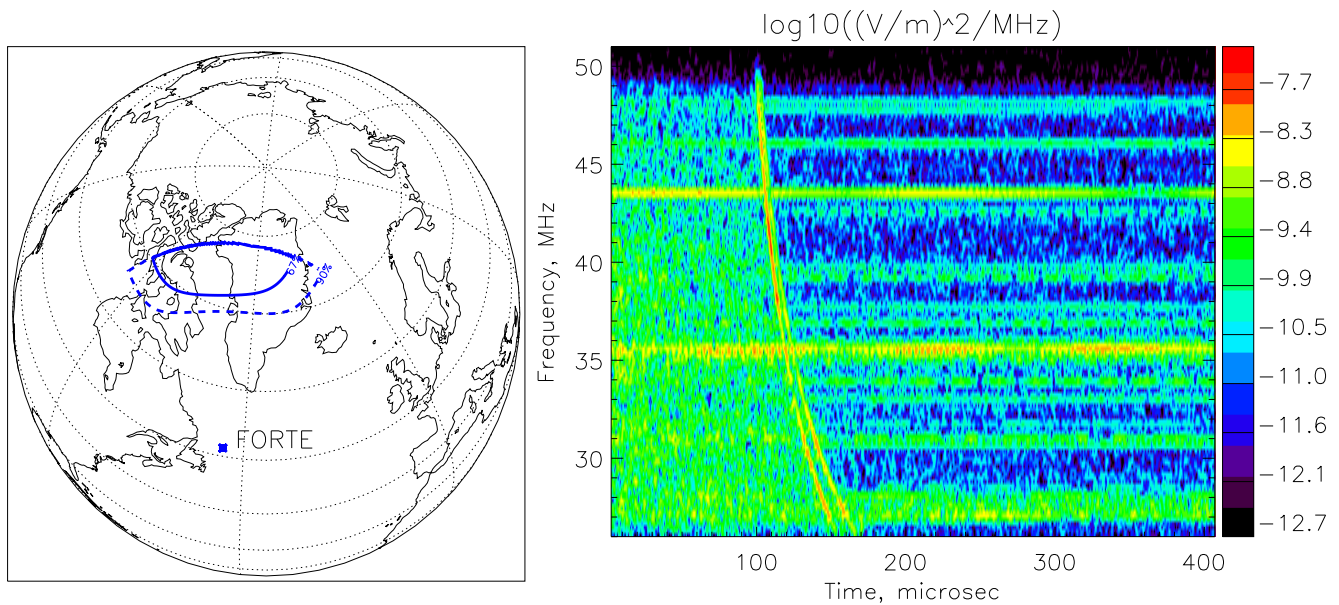}
\caption[splitevent] {\label{fig:splitevent} An example of a
highly-polarized impulsive event detected by FORTE.  The confidence
levels for event location 67\% and 90\% determined using equation
(\ref{eq:bayes}) are shown in the map.  The spectrogram shows the
dispersion of the pulse in the ionosphere and splitting due to Faraday
rotation in geomagnetic field. The horizontal lines are due to
anthropogenic noise (TV and FM radio stations).}
\end{figure*}

The geographic location of the signal source can be determined using the
dispersion of the short electromagnetic pulse in HF range going
through ionosphere. Two important parameters used in
geolocation of the source can be determined from the data from a
single FORTE antenna. The first parameter is the total electron
content (TEC) along the line-of-site between the source and the
satellite. It is proportional to the group time delay, which has
$f^{-2}$ frequency dependence~\cite{Tierney01}.  The second parameter
is determined from the Faraday rotation of a linearly polarized
signal~\cite{Jacobson01}, due to birefringence in magnetoactive
ionospheric plasma. The Faraday rotation frequency turns out to be
equal to the ``parallel'' electron gyrofrequency
$\fcep=eB_\parallel/m_e=f_{ce}\cos\theta$, where $\theta$ is the angle
between the geomagnetic field $\mathbf{B}$ and the ray trajectory at
the intersection with ionosphere.  Both frequency-dependent delay and
frequency splitting due to Faraday rotation are well seen in
Figure~\ref{fig:splitevent}.  The Cherenkov radio emission is expected
to be completely band-limited and linearly polarized, which enables us to
make use of the second parameter for geographic location.

We calculate the probability distribution of the source location using
Bayesian formula:
\begin{equation}
p(\{\lambda,\phi\}|\mathrm{TEC},\fcep)=C
 p(\mathrm{TEC}|\{\lambda,\phi\}) p(\fcep|\{\lambda,\phi\})
\label{eq:bayes}
\end{equation}
where $\{\lambda,\phi\}$ are the latitude and longitude of the source,
$p(\mathrm{TEC}|\{\lambda,\phi\})$ and $p(\fcep|\{\lambda,\phi\})$
are conditional probability distributions for the measured parameters
given the location of the source, and $C$ is a normalization
constant. Here we assume that the measurements of parameters are
independent, and the {\it a priori} distribution of the source
location is uniform in the field of view of the satellite.

To estimate TEC between the source and the satellite (for given
locations of both), we use the Chiu ionosphere model~\cite{Ching73},
adapted to IDL from the FORTRAN source code found at NASA ionospheric
models web site.  This model gives electron density as a function of
altitude for given geographic and geomagnetic coordinates, time of
year, time of day and sunspot number. By integrating it over
altitudes, we find the vertical TEC. To convert it to TEC along the
line-of-sight, we must divide it by the cosine of the angle with the
vertical. Due to the curvature of the Earth, this angle is not
constant along the line-of-sight, and we make an approximation of
taking this angle at the point where the line-of-sight intersects the
maximum of the ionosphere ($F$-layer), at altitude of $\sim$300~km.
The Chiu model, due to simplifying assumptions, does not account for
stochastic day-to-day variability of the vertical electron
content. The standard deviation can be as large as 20--25\% from the
monthly average conditions~\cite[p.\ 10-91]{HandbookGeophys}. Thus, we assume a
Gaussian probability distribution for
$p(\mathrm{TEC}|\{\lambda,\phi\})$ with the center value calculated
using Chiu model and variance of 25\%.

The geomagnetic field is estimated from a simple dipole
model~\cite[pp.\ 4-3, 4-25]{HandbookGeophys}. The error is assumed to
be 10\% according to the estimates for experimental determination of
$\fcep$ from the RF waveform, in Figure~6
of~\cite{Jacobson01}. However, this uncertainty can be greater for
signals that are only partially linearly polarized.  Again, we use a
Gaussian distribution for $p(\fcep|\{\lambda,\phi\})$ with
corresponding central value and the standard deviation of 10\%.

\subsection{Background rejection}
The pulse generated by a UHE $\nu$ shower in ice is expected to be
highly polarized and essentially band-limited up to a few GHz.  In
these aspects, it is similar to the electromagnetic emission from the
``steps'' in a stepped-leader lightning~\cite{Jacobson03}.  However,
the pulses corresponding to lightning steps are accompanied by similar
neighbors before and after, within a time interval from a fraction of
a ms to $\sim$0.5 s.  The signal grouping can thus be used to
distinguish UHE $\nu$ signatures from most lightning events.  Also,
the lightning activity must be present, which is extremely rare in
areas of the Earth covered by ice, and thus can be excluded using the
method described in Section~\ref{ssec:geolocation}.

There is a special type of intracloud lightning which produces isolated
events which are called compact intracloud discharges
(CID)~\cite{Smith98}. However, these events are usually randomly
polarized and have several-$\mu$s pulse durations~\cite{Jacobson03}.

\begin{figure*}
\includegraphics[width=7in]{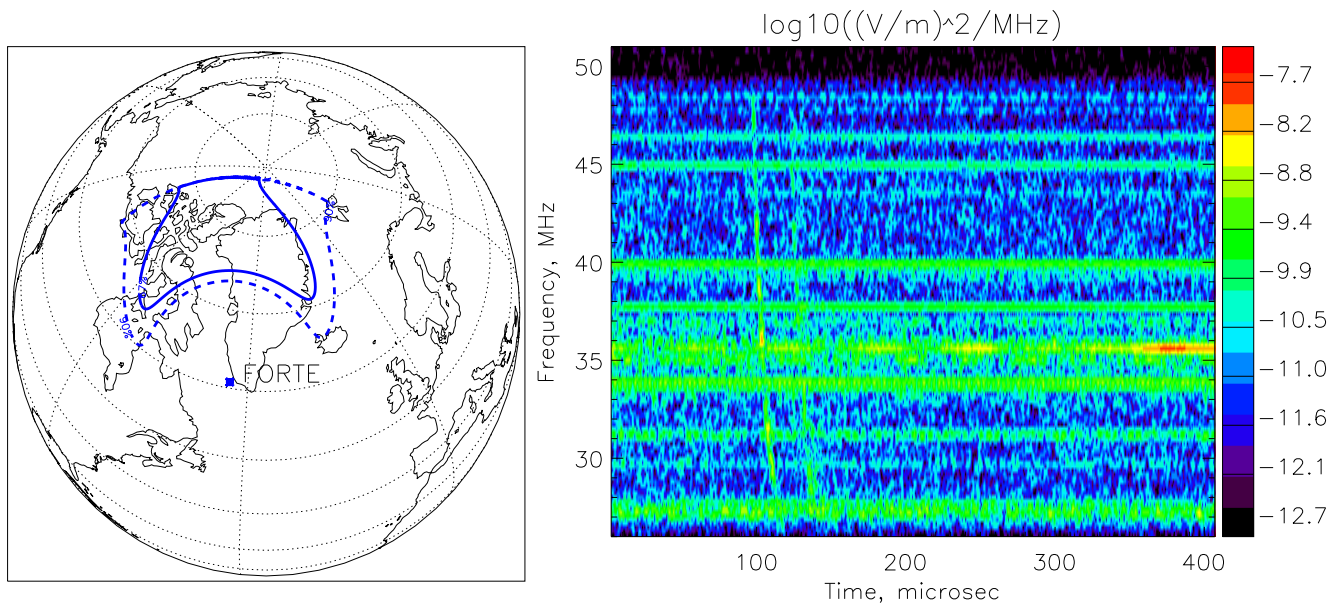}
\caption[tippevent] {\label{fig:tippevent} A TIPP (pulse pair) event
with probable origin location in Greenland but rejected in our
analysis. Notations are the same as in Figure~\ref{fig:splitevent}.}
\end{figure*}

Another rejection method uses the fact that the lightning discharges
occur above ground, and therefore there is a large probability for
FORTE to detect also the signal reflected from the ground. This
phenomenon is known as Trans-Ionospheric Pulse Pairs
(TIPPs)~\cite{Holden95,Massey95,Massey98}. The presence of a second
pulse, therefore, excludes the possibility of the signal to be a UHE
$\nu$ signature. An example of a TIPP event in FORTE data is shown in
Figure~\ref{fig:tippevent}.

\section{FORTE satellite sensitivity to neutrino flux}
\label{sec:sensitivity}

In this section, we estimate the upper limit of the differential flux
of UHE $\nu$ (all flavors) depending on the number of triggers on the
relevant events in the satellite lifetime. The limits are set on the
sum of all flavors since this experiment cannot distinguish between
them.

The typical natural background noise level in FORTE data is
$\sim$10$^{-12}$--10$^{-11}$~(V/m)$^2$/MHz (as can be seen, e.g., from
spectrograms in this paper's Figures).  In a typical 1-MHz trigger
subband this corresponds to RMS value of 1--3~$\mu$V/m.  The trigger
level is set 14--20~dB above the noise, giving the ability to trigger
on impulsive signals with frequency domain values of 5--30~$\mu$V/m in
each 1~MHz trigger subband. For flux limits calculations in this
section, we use the threshold value of 30~$\mu$V\,m$^{-1}$\,MHz$^{-1}$
at $f=38$~MHz, the central frequency of the low FORTE band.

\subsection{Relation between FORTE sensitivity and the limits on
  UHE $\nu$ flux}

Let $\lambda(\En)$ be the theoretical number of triggers of FORTE
satellite in its full lifetime assuming a unit monoenergetic neutrino
flux for different energies $\En$. We will call this function the
sensitivity of the FORTE satellite to UHE $\nu$ flux.  Then the
expected number of triggers is
\[ s=\int \lambda(\En) \Phi(\En) d\En \]
where $\Phi(\En)$ is the differential neutrino flux (per unit area,
time and solid angle).  The number of detected
events is a Poisson distributed random number with expectation $s$.
If no events are detected (null result) we can set a limit on $s$:
\[ s\le\smax=-\log\alpha \]
where $1-\alpha$ is the confidence level. For 90\% confidence level,
for example, we have
\begin{equation}
\int \lambda(\En) \Phi(\En) d\En\le\smax\approx2.3
\label{eq:intlim}
\end{equation}
In general, for $n$ events, the limiting value of $s$ is determined from
\begin{equation}
\alpha=Q(n+1,\smax)
\label{eq:smax}
\end{equation}
where $Q(n+1,\smax)=\frac{\int_{\smax}^\infty x^n e^{-x} dx}{\int_0^\infty
x^n e^{-x} dx}$
is the regularized incomplete gamma function. For example, at a confidence
level of 90\% and $n=1$ we get $s_{\rm up}\approx3.89$.

From the limit set on integrated flux (\ref{eq:intlim}) one can try to
construct a limit on a differential flux. However, such a limit can be
evaded for differential flux models that are anomalous, for example,
if there are very narrow emission lines in the spectrum. Nevertheless, for
reasonable assumptions such as a smooth and continuous model spectrum,
the implied limits are model-independent. As shown in
Appendix~\ref{app:flux}, if we assume that the spectrum is
sufficiently smooth, i.e. does not have any sharp peaks (the peaks
have widths at least of the order of the central energy of a peak),
then we can assert that
\begin{equation}
\Phi(\En)\alt \frac{\smax}{\En\lambda(\En)}
\label{eq:difflim}
\end{equation}
Since this limit does not assume any particular model (except that it
is sufficiently smooth), we will define it as the model-independent limit.

\subsection{Sensitivity calculation}

\begin{figure}
\includegraphics[width=3.375in]{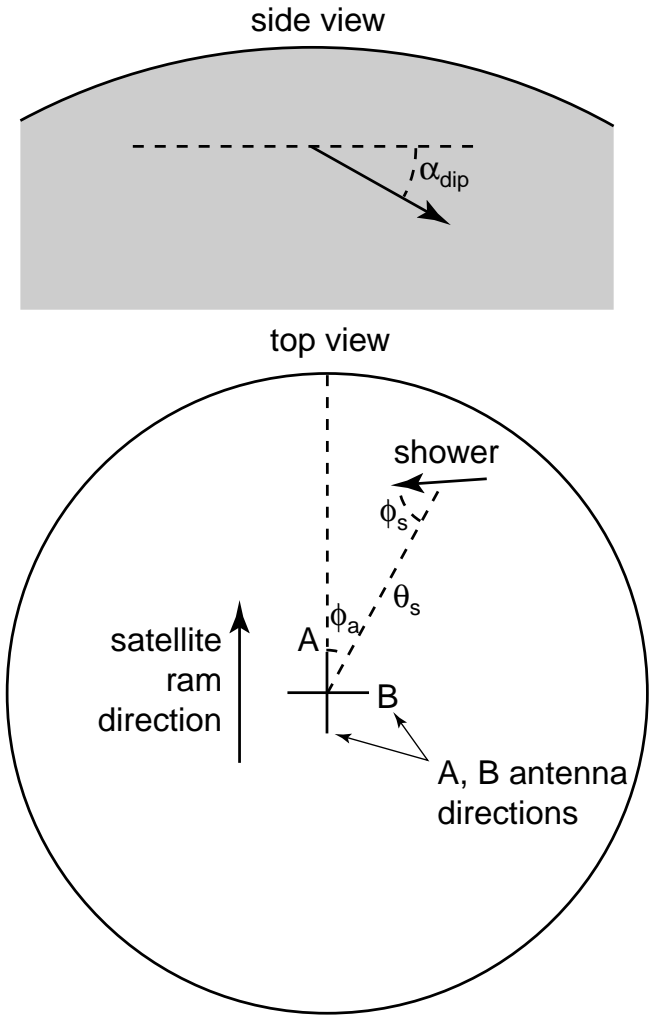}
\caption[view]{\label{fig:view}Configuration for sensitivity calculations.}
\end{figure}

The sensitivity $\lambda(\En_\nu)$, which was introduced in the
previous subsection, is calculated using the specific aperture $R$,
which we define as the trigger rate for unit monoenergetic neutrino
flux when neutrinos interact under a unit area of ice. The specific
aperture $R$ is integrated over the visible ice area averaged over
time (many satellite passes) and multiplied by the entire satellite
lifetime in orbit $T$ (while the two radio receivers of 22~MHz band
were in order) and the fraction of time in trigger mode (duty cycle)
$D$:
\[
\lambda(\Enu)=DT
\left\langle\int_{A_{\rm visible}}
R(\ths,\pha;\Enu)\,dA\right\rangle_{\rm time}
\]
The lifetime of FORTE $T$ is from September 1997 till December
1999. The time when the emission from the Greenland ice sheet is
visible by the satellite can be estimated by calculating the total
time spent by FORTE inside a circle of radius $20^\circ$ arc distance
from the approximate geographic center of Greenland, at $70^\circ$N,
$40^\circ$W. This time is estimated to be $\sim$38~days. The duty
cycle $D$ is estimated to be 6\% by calculating the run time fraction
within this circle, making the effective time of observation to be
$\sim$3~days.

The specific aperture $R$ is a function of the position of the
neutrino-generated shower which is given by the arc distance to the
shower location $\ths$ and its azimuthal angle $\pha$ (see
Figure~\ref{fig:view}). Note that the visible area $A_{\rm visible}$
depends on the satellite location. Averaging over time in our
calculations is replaced by averaging over satellite position over the
Earth surface with a weight corresponding to the fraction of time the
satellite spends in a given point of the globe.

To calculate the specific aperture $R$, we must calculate how many
neutrinos interact in ice, depending on energy $\Enu$, dip angle
$\dipang$ (the angle of $\nu$ velocity below the horizon at the
interaction point) and the depth of interaction $z$.  We use the
theoretical neutrino-nucleon interaction cross-section $\sigma_{\nu
N}(\Enu)$~\cite{Kwiecinski99}.  The neutrino flux from below is greatly
reduced by interactions in the Earth volume, and most detectable
interactions are from horizontal and down-going events. Here we have to
note that this statement does not apply to neutralino
interactions since neutralino-nucleon interaction cross-section
$\sigma_{\chi N}$ can be much smaller than neutrino cross-section at
a similar energy~\cite{neutralino}.

The number of interactions in ice is characterized by function
$\fluxfraction(\Omega,z;\Enu)$ defined in the following way:
\[
\Phi\fluxfraction(\Omega,z;\Enu)\,d\Omega\,dV=N_{\rm nuc}
\sigma_{\nu N}(\Enu)\,d\Phi\,dV
\]
Here $\Omega=(\dipang,\phs)$ is the neutrino velocity direction (see
Figure~\ref{fig:view}), $z$ is the interaction depth, $N_{\rm nuc}$ is
the nucleon number density and $d\Phi=\Phi e^{-\tau} d\Omega$ is the
flux in solid angle element $d\Omega=\sin\dipang d\dipang d\phs$
attenuated by neutrino absorption in a layer of optical thickness
$\tau(\dipang,z)=\int N_{\rm nuc}\sigma_{\nu N}dl$.

The specific aperture $R$ is found using the fraction of interacting
neutrinos which produce field $E_{\rm ant}>E_{\rm th}$ at the
satellite antennas:
\begin{eqnarray*}
\lefteqn{R(\ths,\pha;\Enu)=\int \fluxfraction(\Omega,z;\Enu)}\\
& &\left(
\int_0^1 \Theta[E_{\rm ant}(y\Enu,\Omega,\ths,\pha)-E_{\rm th}] p(y)\,dy
\right)\,d\Omega\,dz
\end{eqnarray*}
where $\Theta$ is the step function, and $p(y)=(1/\sigma)(d\sigma/dy)$
is the probability distribution function for the kinematic parameter
$y=\En_{\rm had}/\Enu$, the fraction of energy going into the hadronic
shower. The $p(y)$ dependence is taken from~\cite{Gandhi96} for the
highest energy considered in that paper, $\Enu=10^{12}$~GeV.  Since
$E_{\rm ant}(y\Enu,\Omega,\ths,\pha)= yE_{\rm
ant}(\Enu,\Omega,\ths,\pha)$, we can integrate over $y$ immediately
and get
\begin{eqnarray*}
\lefteqn{R(\ths,\pha;\Enu)=\int \fluxfraction(\Omega,z;\Enu)}\\
& &\left[1-F_y\left(
\frac{E_{\rm th}}{E_{\rm ant}(\Enu,\Omega,\ths,\pha)}
\right)\right] d\Omega dz
\end{eqnarray*}
where $F_y(y)=\int_0^y p(y') dy'$ is the cumulative distribution
function of $y$.

\begin{figure*}
\includegraphics[width=7in]{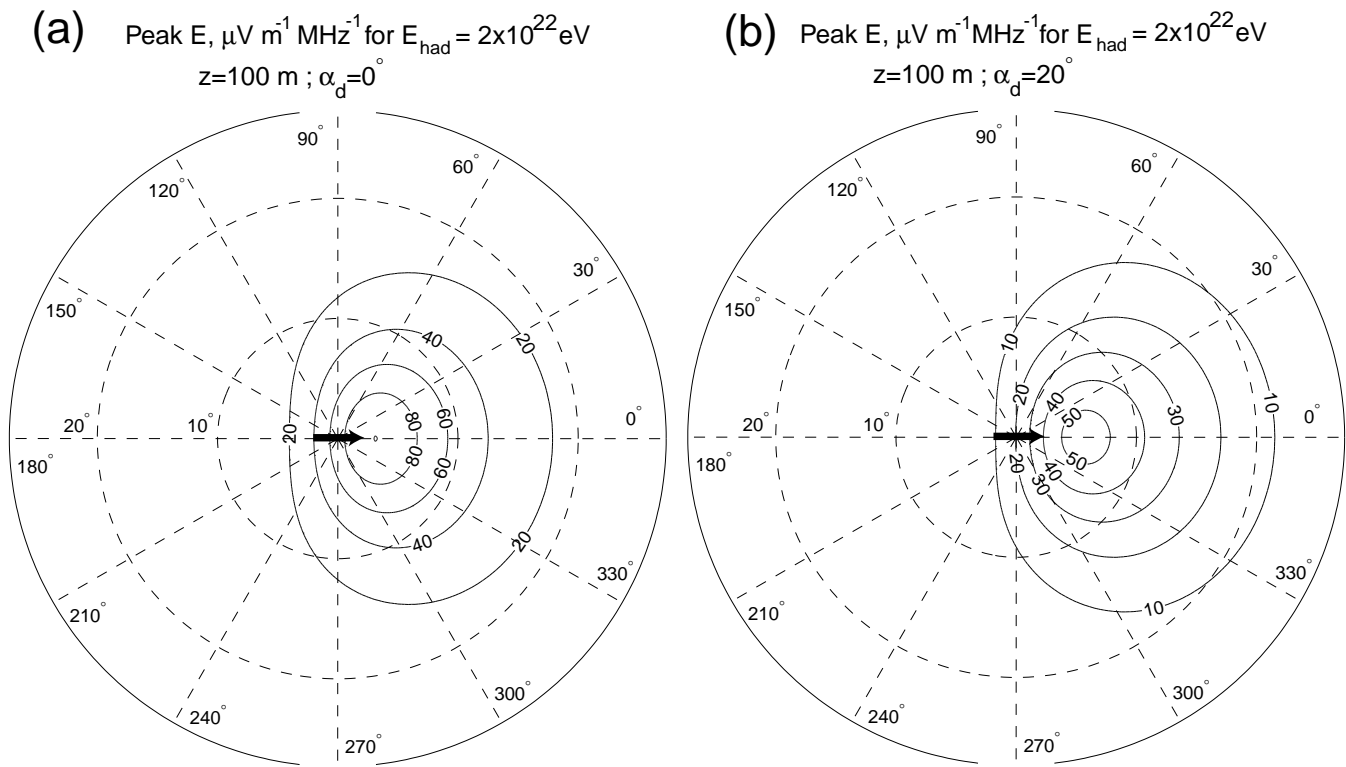}
\caption[efield]{\label{fig:efield} The peak electric field at the
satellite altitude $E_{\rm sat}$, emitted in a Cherenkov process by a
hadronic shower of energy $\sci{2}{22}$~eV. These plots do not include
the antenna response. The shower is in the center of the circle, and
is directed to the right. Emission is shown for interaction at depth
$z=100$~m and dip angles of (a) $\dipang=0^\circ$ and (b)
$\dipang=20^\circ$. The dashed circles represent the arc distance in
degrees ($1^\circ\approx111$~km on the Earth surface).}
\end{figure*}

Given a neutrino interacting at dip angle $\dipang$ and depth $z$ in
ice, we can calculate the field detected at the satellite.  First, the
emitted field is given by Cherenkov emission formula
(\ref{eq:cher}). Then the field is attenuated in ice, refracted at the
ice surface and detected at the satellite taking into account the
directional antenna response.  In the subsections of
Appendix~\ref{app:efield}, we give the details of calculations for
these steps. In Figure~\ref{fig:efield}, we show the peak electric
field at the satellite altitude (800~km) at the central frequency of
the low FORTE band ($f=38$~MHz). As we see, even for a dip angle of
$20^\circ$ there is a significant emission upward from the shower due
to the width of Cherenkov cone. Although the field plotted in
Figure~\ref{fig:efield} does not include the antenna response, the
plots can give some idea about where the satellite has to be to see
the signal at a given threshold, the typical value of the threshold being
30~$\mu$V\,m$^{-1}$\,MHz$^{-1}$.

\section{Results}
\label{sec:results}

\subsection{Event search results}

We searched for events recorded while FORTE was inside a circle of
radius of 20$\degree$ with a center at 70$\degree$N, 40$\degree$W, in
time period from the start of FORTE in September 1997 to December
1999, when both 22-MHz-bandwidth receivers were
lost~\cite{RousselDupre01}. We estimate that the satellite spent a
total of 38 days inside this circle, with at least $\sim$6\% of it
being the time in trigger mode. We found a total of 2523 events. From
these, only 77 are highly polarized.  These 77 events can be
geolocated using both parameters described in
subsection~\ref{ssec:geolocation}. Of these, only 16 events have
intersection of the 90\% confidence level with Greenland's ice
sheet. Out of the remaining 16 events, 11 are rejected for being TIPP
events, i.e. pulse pairs with ground reflections that indicate that
the origin locations are above ground. An example of a rejected TIPP
event is shown in Figure~\ref{fig:tippevent}. Two more events were
rejected because of the presence of a precursor before the pulse,
which is characteristic of certain type of lightning~\cite{Jacobson02}
and cannot be present in a neutrino shower signal. An example of such
event was shown in Figure~\ref{fig:splitevent}.

\begin{figure*}
\begin{center}
\begin{tabular}{c}
\includegraphics[width=7in]{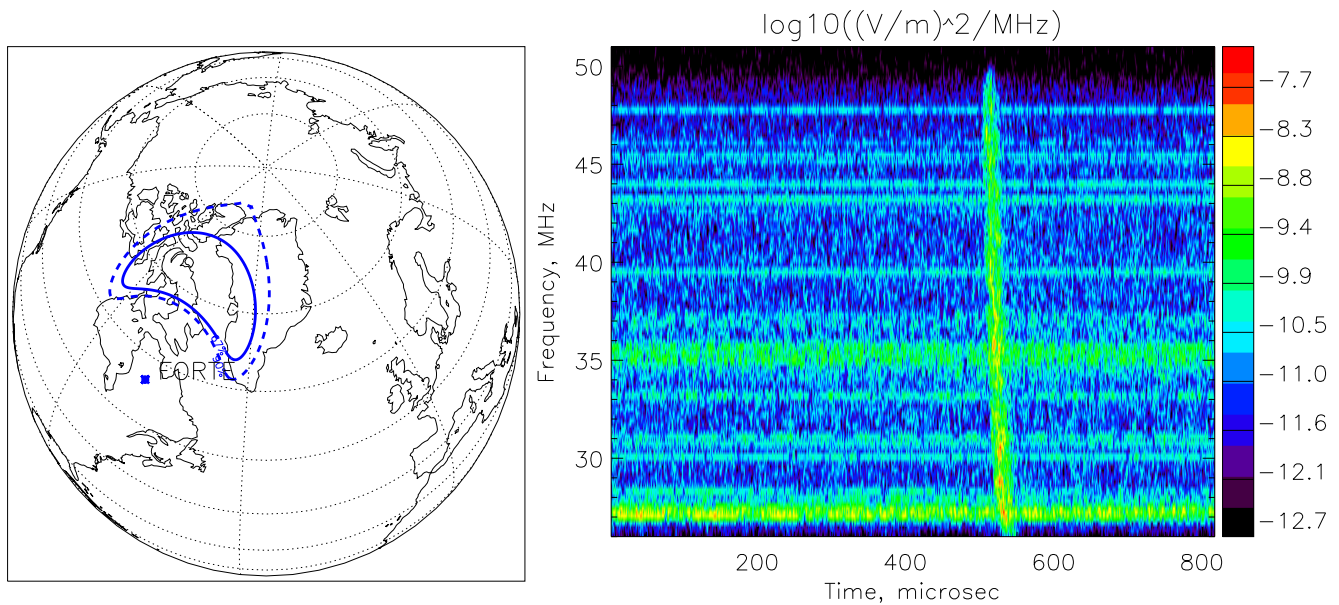}
\end{tabular}
\end{center}
\caption[goodevent] {\label{fig:longevent} The confidence levels of
67\% and 90\% for geographic location and a spectrogram of an example
event.  This event cannot be generated by a neutrino due to its long
duration ($\agt$10~$\mu$s).  }
\end{figure*}

\begin{figure*}
\begin{center}
\begin{tabular}{c}
\includegraphics[width=7in]{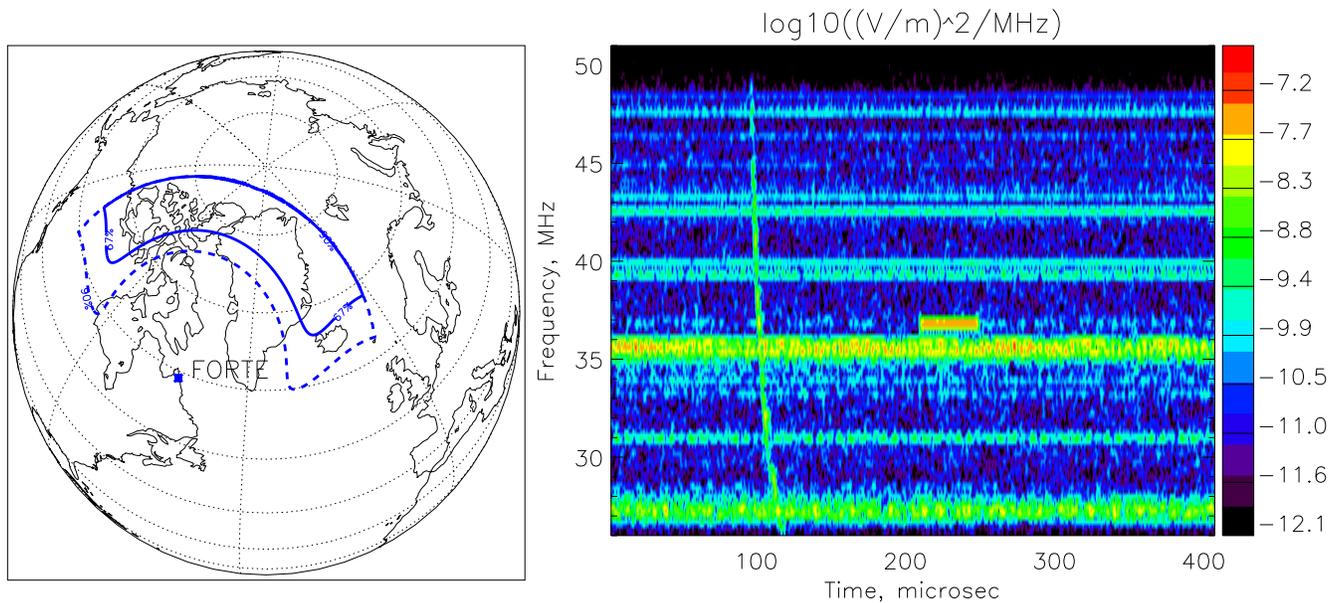}
\end{tabular}
\end{center}
\caption[goodevent] {\label{fig:goodevent} The confidence levels of
67\% and 90\% for geographic location and a spectrogram of an example
event.  A short horizontal streak in the spectrogram is due to
anthropogenic noise (a radar).  This event needs further consideration
for being rejected as neutrino-generated. However, its nearest
neighbor events were found at $-1.4$~ms and $+0.7$~ms, which makes it
a probable lightning event.  }
\end{figure*}

\begin{figure*}
\begin{center}
\begin{tabular}{c}
\includegraphics[width=7in]{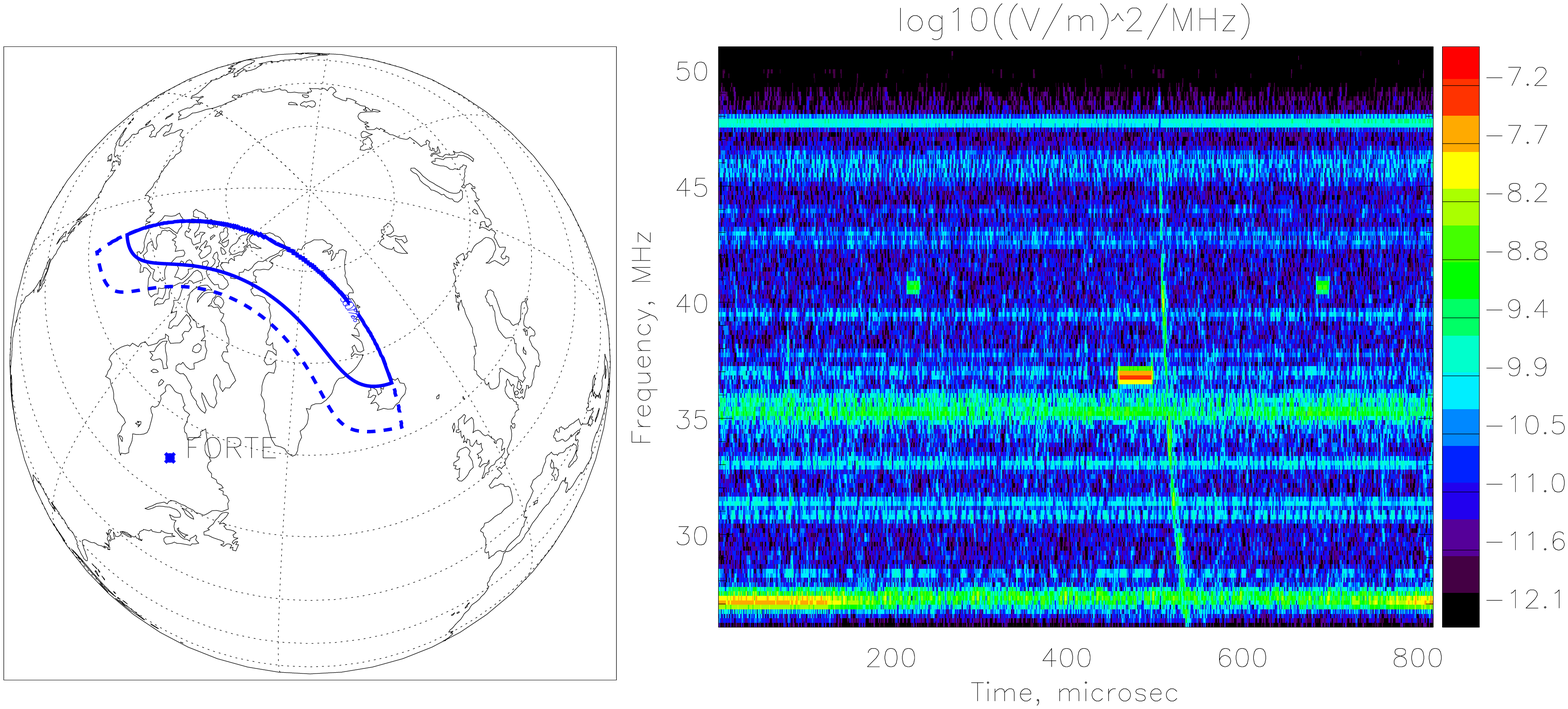}
\end{tabular}
\end{center}
\caption[goodevent] {\label{fig:goodevent2} The confidence levels of
67\% and 90\% for geographic location and a spectrogram of an example
event.  Its nearest neighbors were found at $-0.27$~s and $+5.55$~s.
This event needs further consideration for being rejected as
neutrino-generated.  }
\end{figure*}

Out of the remaining three events, one (shown in
Figure~\ref{fig:longevent}) is rejected for its long duration
($\agt$10~$\mu$s), since the Cherenkov pulse is expected to be only
$\sim$1~ns long (the time resolution of the FORTE detector limits this
to $\agt$20~ns). The remaining two events are shown in
Figures~\ref{fig:goodevent} and \ref{fig:goodevent2}. The first of
them has close neighbor events (at $-1.4$~ms and $+0.7$~ms), which
makes it a probable part of a stepped-leader process in a
lightning. The neighbors of the second event are not very close (at
$-0.27$~s and $+5.55$~s), but it still can be a lightning event. The
recent analysis by the FORTE team~\cite{Jacobson03} has shown that the
lightning events are likely to have neighbors in $\pm0.5$~s interval,
with the most probable separation of $\pm0.01$~s, and the accidental
coincidence rate is $\sim$0.9~per second. This value of the accidental
coincidence rate makes the candidate event shown in
Figure~\ref{fig:goodevent2} indistinguishable from an isolated
lightning discharge.  The analysis of this event continues.

\subsection{Flux limits set by FORTE}

\begin{figure}
\includegraphics[width=3.375in]{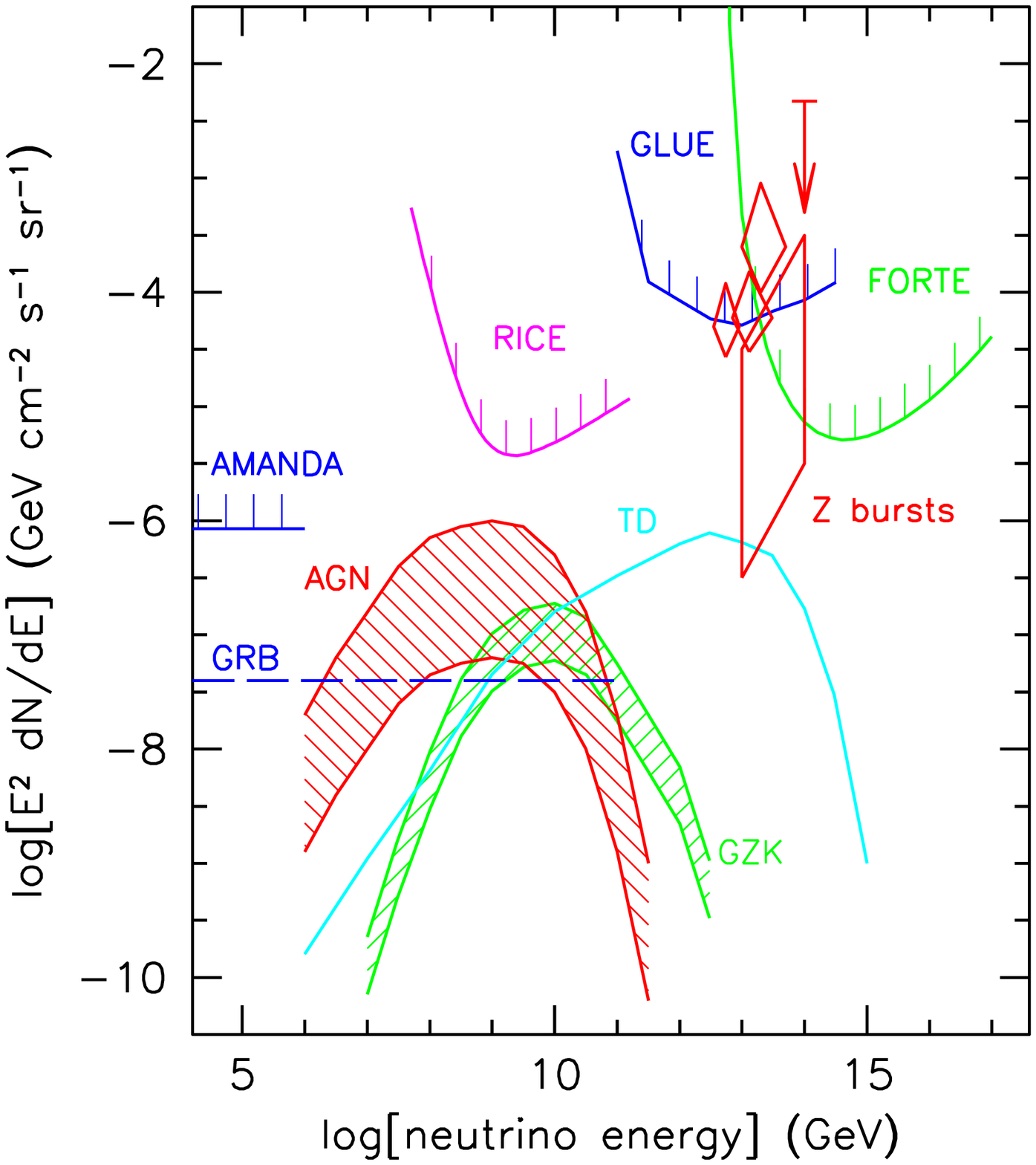}
\caption[fluxlimits]{\label{fig:fluxlimits}The estimated limits on UHE
$\nu$ flux detectable by FORTE using Greenland ice sheet. The limit is
compared to predicted neutrino fluxes from various sources: GRB: gamma
ray bursts~\cite{GRB}; AGN: active galactic nuclei~\cite{AGN}; GZK:
neutrinos produced in GZK mechanism from cosmic rays~\cite{GZKnu}; TD:
topological defect model (non-SUSY)~\cite{TDSigl}; Z burst models are:
the slanted box by~\cite{Weiler01}, the diamonds showing the 1$\sigma$
level errors are by~\cite{Fodor02b,Ringwald03}, and the arrow is
by~\cite[default value of parameters]{Gelmini00}. The limits from
other experiments are also shown: AMANDA~\cite{AMANDAlim},
RICE~\cite{RICE4} (determined using (\ref{eq:difflim}) from the
effective volume and time of observation),
GLUE~\cite{GLUE1,GLUE2}. The limits from all experiments show the
limits on the combined flux of neutrinos of all flavors (assumed mixed
in equal amounts), except AMANDA which is only sensitive to $\nu_\mu$
($\bar{\nu}_\mu$).}
\end{figure}

In Table~\ref{tab:table2} we give the calculated FORTE neutrino flux
sensitivity values $\lambda(\En_\nu)$. On the basis of these values
one can set a limit on any model flux $\Phi(\En_\nu)$ using numerical
integration in equation~(\ref{eq:intlim}), with $\smax=3.89$ since we
have one uncertain event as our background noise.  In
Figure~\ref{fig:fluxlimits} we plot the model-independent flux limits
set by equation~(\ref{eq:difflim}) on the basis of these data (again
with $\smax=3.89$).  In the same Figure we also show the comparison of
calculated flux limits with predicted neutrino fluxes from various
sources.  Some of the sources of super-GZK neutrinos are reviewed,
e.g. in~\cite{Berezinsky03}.  As we see, the flux limits set by FORTE
observations of the Greenland ice sheet can reject some regions of
parameters of the Z burst
model~\cite{Fargion99,Weiler99,Weiler01,Gelmini00,Fodor02a,Fodor02b}.

\begin{table}
\caption{\label{tab:table2}The FORTE sensitivity
$\lambda(\En)$~(cm$^2$\,s\,sr) to neutrino flux (any neutrino flavor)
and neutralino flux (for different neutralino-nucleon
cross-sections).}
\begin{ruledtabular}
\begin{tabular}{ccccc}
$\log_{10}\En$ & $\lambda(\En_\nu)$ & $\lambda(\En_\chi)$
  & $\lambda(\En_\chi)$ & $\lambda(\En_\chi)$ \\
(GeV) &  & $\sigma_{\nu N}$ & $0.1\sigma_{\nu N}$ & $0.01\sigma_{\nu N}$ \\
\hline
13.0 & $\sci{8.0}{12}$ & $\sci{2.1}{14}$ & $\sci{2.6}{13}$ & $\sci{4.1}{12}$ \\
13.5 & $\sci{5.4}{14}$ & $\sci{5.6}{15}$ & $\sci{6.6}{14}$ & $\sci{7.9}{13}$ \\
14.0 & $\sci{5.3}{15}$ & $\sci{3.2}{16}$ & $\sci{3.8}{15}$ & $\sci{4.3}{14}$ \\
14.5 & $\sci{2.4}{16}$ & $\sci{9.7}{16}$ & $\sci{1.2}{16}$ & $\sci{1.3}{15}$ \\
15.0 & $\sci{7.1}{16}$ & $\sci{2.2}{17}$ & $\sci{2.8}{16}$ & $\sci{3.0}{15}$ \\
15.5 & $\sci{1.7}{17}$ & $\sci{4.1}{17}$ & $\sci{5.6}{16}$ & $\sci{6.1}{15}$ \\
16.0 & $\sci{3.4}{17}$ & $\sci{6.9}{17}$ & $\sci{1.0}{17}$ & $\sci{1.1}{16}$ \\
16.5 & $\sci{6.0}{17}$ & $\sci{1.0}{18}$ & $\sci{1.7}{17}$ & $\sci{1.9}{16}$ \\
17.0 & $\sci{9.5}{17}$ & $\sci{1.4}{18}$ & $\sci{2.6}{17}$ & $\sci{3.0}{16}$ \\
\end{tabular}
\end{ruledtabular}
\end{table}

\begin{figure}
\includegraphics[width=3.375in]{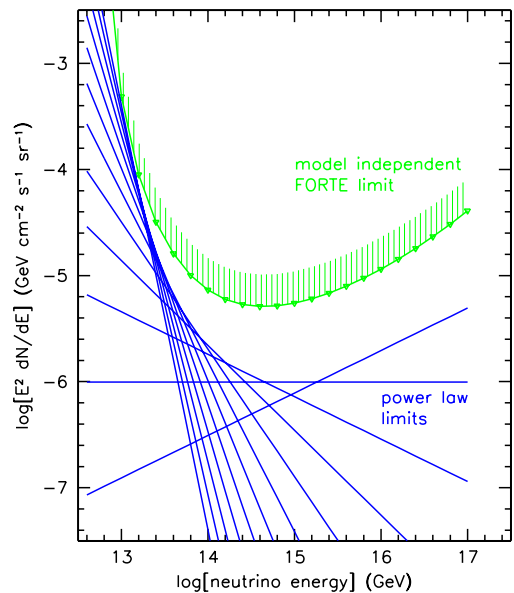}
\caption[powerlaws]{\label{fig:powerlaws}Upper curve: the differential
flux limit of equation (\ref{eq:difflim}); lower set of lines: the
limits set by equation (\ref{eq:bestdifflim}) assuming power law flux
shapes $K(\En;\alpha)=\En^{-\alpha}$ for $\alpha$ in the interval from
$1.6$ to $6.0$ with a step of $0.4$.}
\end{figure}

Note that differential fluxes in some models can be even smoother and
wider in energies than assumed for derivation of
(\ref{eq:difflim}). Figure~\ref{fig:powerlaws} shows what would be the
limits on a class of models with power-law energy dependence of the
flux, $\Phi(\En)\propto\En^{-\alpha}$. A similar analysis using
power-law models was performed in the past, e.g., for RICE
detector~\cite{RICE2}.

\begin{figure}
\includegraphics[width=3.375in]{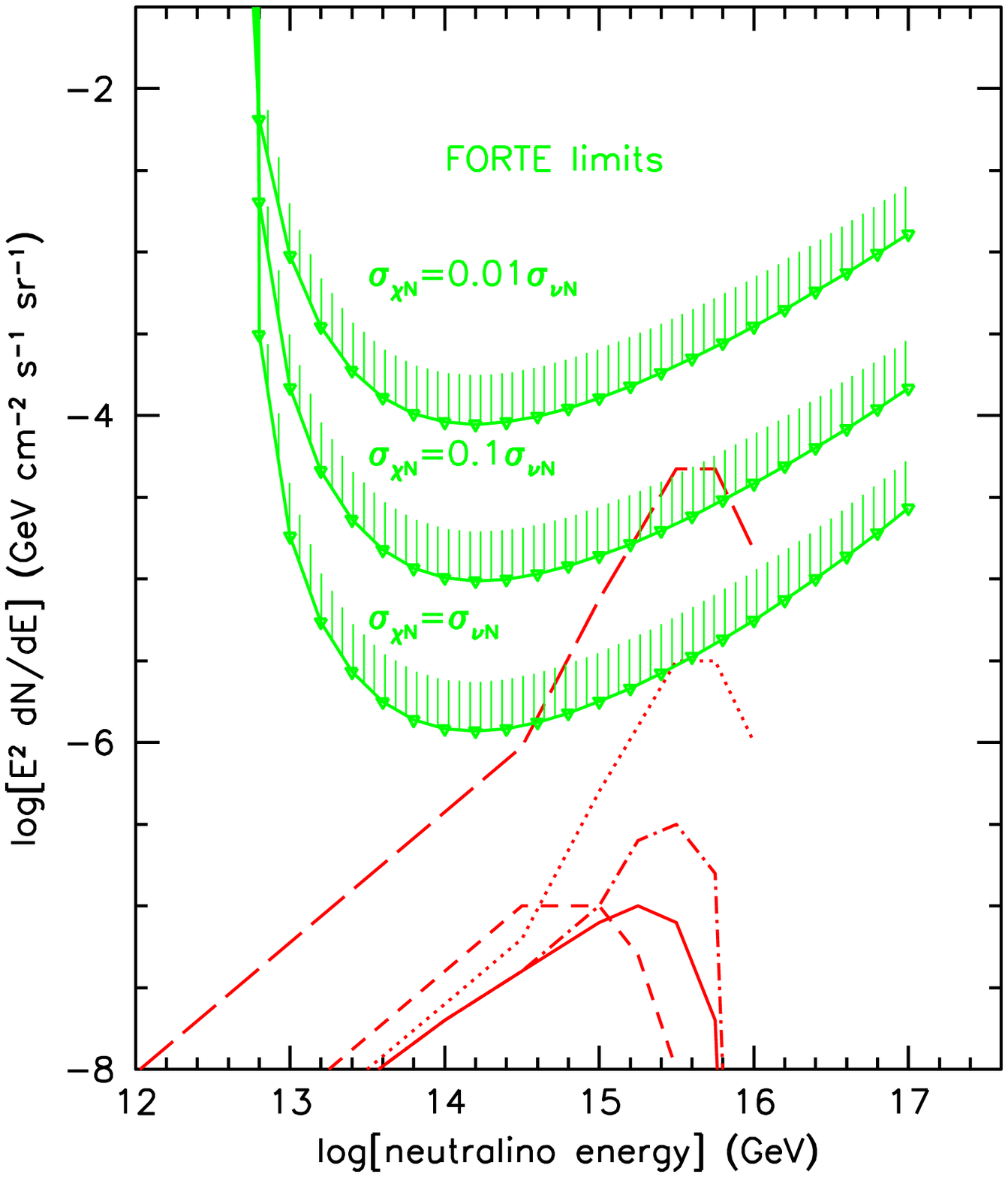}
\caption[neutralino]{\label{fig:neutralino}The limits on neutralino
fluxes set by FORTE observations of the Greenland ice sheet, for one
detected event and different assumptions about neutralino-nucleon
cross-sections. Shown also are predicted neutralino
fluxes~\cite{neutralino} for decay of superheavy particles of
$M_X=\sci{2}{25}$~eV. The lower four curves are for $X$ decays in the
halo of the Galaxy, with the primary decay into: (1)~quark$+$antiquark
(solid); (2)~quark$+$squark (dot-dash); (3)~$SU(2)$ doublet
lepton$+$slepton (dots); (4)~5 quark$+$5 squark (short dashes).  The upper
curve (long dashes) is for homogeneous $X$ distribution (in which case
the flux is enhanced by a factor of 15 compared to a ``galactic''
distribution), decay scenario 3. See also Table~\ref{tab:table1}.}
\end{figure}

As another result of present research, we set limits on the flux of
neutralinos, weakly interacting particles predicted by the Minimal
Supersymmetric Standard Model (MSSM). The calculations are performed
in the same way as for setting the neutrino flux limits with a few
differences. First, the cross-section for nucleon interaction
$\sigma_{\chi N}$ is different, and is expected to be in the range
from $\sim\!(1/100)\sigma_{\nu N}$ to $\sim\!\sigma_{\nu
N}$~\cite{neutralino}.  Second, all of the energy goes into the
hadronic shower. The results are presented also in Table~\ref{tab:table2}.
In Figure~\ref{fig:neutralino}, we compare the
predicted neutralino fluxes~\cite{neutralino} with the
model-independent flux limits set by FORTE. We see that for
neutralino-nucleon cross-sections in the range $\sigma_{\chi N}\agt
0.1\sigma_{\nu N}$, strong limits are set on the model predicting
neutralino flux from decay of heavy $X$ particles with
$M_X=\sci{2}{25}$~eV, especially if the sources are homogeneously
distributed. Note that for a case of a different mass,
$M_X=\sci{2}{21}$~eV, which is also considered in~\cite{neutralino},
we are unable to set any limits since all decay products are below
FORTE threshold.

In Table~\ref{tab:table1}, we apply the more robust model-dependent
limit to these models to get the confidence level of rejection
according to equation (\ref{eq:smax}), for $n=1$ (one uncertain
event).  We vary cross-section $\sigma_{\chi N}$, take different decay
scenarios and $X$ distributions. As one can see from this Table,
several models are rejected with very high confidence.

\begin{table*}
\caption{\label{tab:table1}The confidence levels for rejecting the
models of neutralino production by heavy $X$ particle decay with
$M_X=\sci{2}{25}$~eV~\cite{neutralino}. We only show models with the
rejection confidence level $>$50\%. The scenarios of $X$ decay are the
same as in Figure~\ref{fig:neutralino}. The variable $\alpha$ is
defined as $1-\textrm{CL}$.}
\begin{ruledtabular}
\begin{tabular}{cccccc}
$\sigma_{\chi N}/\sigma_{\nu N}$  & decay scenario & $X$ distribution
  & expected number of triggers &
  $\alpha$  & CL\\
\hline
1 & 1 & homogeneous & 8.2927 & $\sci{2.3264}{-3}$ & 99.7674\% \\
1 & 2 & homogeneous & 14.0678 & $\sci{1.1708}{-5}$ & 99.9988\% \\
1 & 3 & halo & 6.7491 & $\sci{9.0814}{-3}$ & 99.0919\% \\
1 & 3 & homogeneous & 101.2365 & $\sci{1.1044}{-42}$ & 100\% \\
1 & 4 & homogeneous & 11.6834 & $\sci{1.0696}{-4}$ & 99.9893\% \\
0.1 & 2 & homogeneous & 1.8251 & $\sci{4.554}{-1}$ & 54.4602\% \\
0.1 & 3 & homogeneous & 13.6869 & $\sci{1.6702}{-5}$ & 99.9983\% \\
\end{tabular}
\end{ruledtabular}
\end{table*}

\section{Discussion}

The limits shown in Figures~\ref{fig:fluxlimits}
and~\ref{fig:neutralino} represent to our knowledge the first direct
experimental limits on the fluxes of neutrinos and other weakly
interacting particles in this energy range. The fact that the first
such limits already have constrained several proposed models is an
indication of the power of the radio detection techniques, but the
scarcity of other limits in this regime also suggests that they be
accepted with caution. Here we discuss briefly some of the potential
issues with these constraints.

  At the energies to which FORTE is sensitive, the energy of the pulsed 
coherent radio emission can become one of the dominant energy-loss 
mechanisms for the shower. This implies that the radiation reaction of 
the shower to the pulse could lead to modification of the shower 
development, and consequently some change in the radiation parameters. 
We have not attempted to correct for this effect in our analysis, but we 
note that it is probably not important below $\sim 
10^{24}$~eV~\cite{AccelAsk1}. Above this energy we expect that the shower 
radiation pattern might spread to some degree, depending on the 
foreshortening of the cascade length due to radiation reaction 
deceleration.

  We have noted that our limits extend into the mass scale for GUT
particles. However, it is important to note that the
center-of-momentum energies for interactions of these neutrinos on any
other standard particles are of order 10~PeV or less. Our results thus
depend on a $\leq$3 order of magnitude extrapolation of the standard
model neutrino cross sections from the current highest energy estimate
from accelerators at $\sim$30~TeV~\cite{HERA1,HERA2,HERA3}, over an
energy regime where the cross sections grow only logarithmically with
energy. For this reason we do not expect that the energy scale itself
is good cause to doubt the values for the flux limits.

  At 10~PeV center-of-momentum energies, interactions of the primary 
neutrinos will exceed in CM energy those of any observed ultra-high 
energy cosmic rays, and could therefore lead to production of new heavy 
particles in the showers themselves. Such interactions could include 
channels in which most of the energy goes into an unobservable particle, 
or a particle with interactions much weaker than neutrinos. In the 
absence of any specific proposals for models and interactions, we can 
only note that such behavior can evade our limit, but could lead to 
other observable secondary particles with different angular 
distributions that we have not considered.

Since FORTE threshold for detection of a weakly interacting particles
($\sim\!10^{22}$~eV) is higher than the GZK limit
($\sim\!\sci{5}{19}$~eV), it sets the limits on neutrinos producing
super-GZK cosmic rays through resonant interaction with background
neutrinos within $\sim$50~Mpc distance from us, the Z burst
mechanism~\cite{Fargion99,Weiler99,Weiler01,Gelmini00,Fodor02a,Fodor02b}.
Although FORTE sets limits on parameters of Z burst models, the
uncertainties in the models and the measured super-GZK cosmic ray flux
still make most Z burst scenarios consistent with FORTE data.

The strong FORTE constraints on the neutralino production
model~\cite{neutralino} from heavy $X$ particles, sets a joint limit
on (1)~$X$ particle distribution and mass (2)~$X$ particle decay
channels and (3)~neutralino-nucleon cross-section. Since some of $X$
particle decay scenarios (at the mass $\sci{2}{25}$~eV) are strongly
rejected, as shown in Figure~\ref{fig:neutralino} and
Table~\ref{tab:table1}, this can give an insight into the possible
nature of such particles and the physics at supersymmetric
grand-unification scale. Although reference~\cite{neutralino} does not
consider models with masses intermediate between $\sci{2}{21}$ and
$\sci{2}{25}$ eV, we expect that FORTE's sensitivity, which extends an
order of magnitude or more below the region where the model at
$\sci{2}{25}$ is constrained, will also limit neutralinos of masses
down to $\sim\!10^{24}$~eV, particularly if the cross sections
approach $\sigma_{\nu N}$.

\section{Conclusions}
We have performed a search for radio frequency signatures of
ultra-high energy neutrinos originating from coherent Cherenkov
emission from cascades in the Greenland ice sheet, observed with the
FORTE satellite over an $\sim$2~year period. In $\sim$3~days of net
exposure, a single candidate, presumed to be background, survives the
analysis, and we set the first experimental
limits on neutrino fluxes in the $10^{22}$--$10^{25}$~eV energy
region. These limits constrain the available parameter space for the Z
burst model. In addition we constrain several variations of a model
which involves light super-summetric particles (neutralinos) at these
energies, particularly those with interaction cross-sections
approaching those of neutrinos.

\appendix

\section{Cherenkov emission from electromagnetic showers}
\label{app:cherenkov}
Let us model the shower as a point charge moving with the speed of
light. Then the current is given by
$J_z(\mathbf{r},t)=cq(z)\delta(\mathbf{r}-c\hat{z}t)$ and $J_x=J_y=0$.
The Fourier transform is
\[
J_z(\mathbf{r},\omega)
=2\int J_z e^{i\omega t}dt
=2q(z)\delta(x)\delta(y)e^{i\omega z/c}
\]
(The factor of 2 is for consistency with our definition of $E(\omega)$)
The frequency-domain vector potential $\mathbf{A}$ satisfies Helmholtz
equation $\nabla^2A_z+k^2A=-\mu\mu_0J_z$ and $A_x=A_y=0$, where
$k=n\omega/c$ and $n=\sqrt{\epsilon\mu}$.  Its solution at the
observation point $\mathbf{R}$ is
\[
A_z(\mathbf{R})=\mu\mu_0\int
\frac{e^{ikR'}}{4\pi R'} J_z(\mathbf{r}) d^3 \mathbf{r}
\]
where $R'=|\mathbf{R}-\mathbf{r}|$.
In the Fraunhofer zone, the standard approximation is
\[
\frac{e^{ikR'}}{4\pi R'}\approx \frac{e^{ikR}}{4\pi R}
e^{-i(\mathbf{k}\cdot\mathbf{r})}
\]
Thus,
\[
A_z(\mathbf{R},\omega)=\mu\mu_0\frac{e^{ikR}}{2\pi R}\int_{-\infty}^{+\infty}
q(z)e^{-iz(k\cos\theta-\omega/c)}dz
\]
where $\theta$ is the emission angle.

The magnetic induction is $\mathbf{B}=\nabla\times\mathbf{A}$. In
the far zone $B=|\mathbf{B}|=kA_z\sin\theta$ and the electric field is
$E=cB/n=\omega A_z\sin\theta$.

Let us consider a Gaussian shower profile
$q(z)=Qe^{-\frac{z^2}{2L^2}}$, where $Q$ is the maximum attained
charge excess and $L$ is the characteristic shower length. Then
\[
R|\E(\omega)|=\frac{\mu\mu_0 Q L\omega}{\sqrt{2\pi}}
 \sin\theta e^{-(kL)^2(\cos\theta-1/n)^2/2}
\]

\section{Cherenkov emission from LPM showers}
\label{app:lpm}
\begin{figure*}
\includegraphics[width=7in]{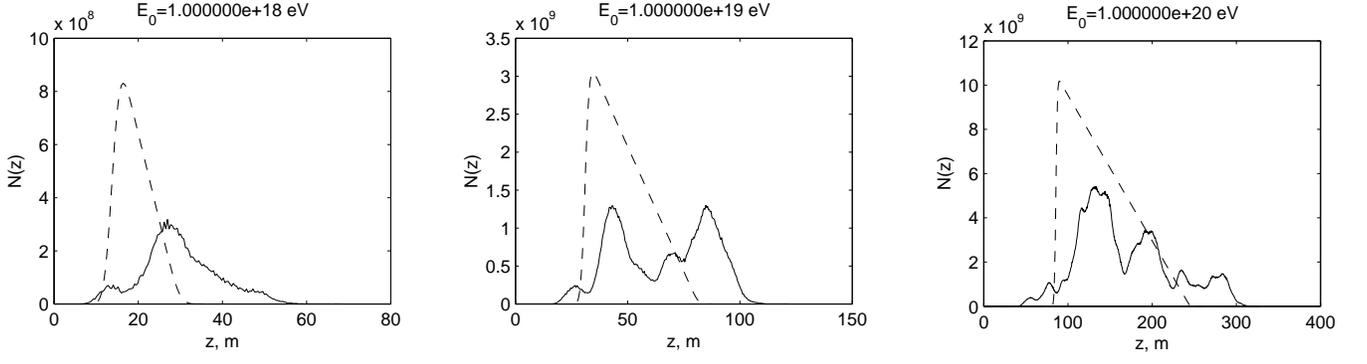}
\caption[lpmmodel]{\label{fig:lpmmodel}
Results of Monte Carlo simulation compared to simple
analytical model presented here.}
\end{figure*}

According to \cite{Stanev82}, LPM effect is important for
particle energies $\En>\Elpm$, where $\Elpm=61.5 X_0$~TeV, where
$X_0$ is the radiation length in cm. The radiation length in
mass units is 36.1 g/cm$^2$ in water,
giving $X_0=39.1$~cm in ice since $\rho_{\rm
  ice}=0.924$~g\,cm$^{-3}$. Thus, $\Elpm=2.4$~PeV. The increased radiation
length for bremsstrahlung and the 4/3 of the mean free path  for pair
production, according to the same paper, are given approximately
(within $\sim$20\%) by $X_{\rm LPM}=\sqrt{\En/\Elpm}X_0$, when
$\En\gg\Elpm$. Let us model the UHE electromagnetic shower as the
initial particle gradually losing its energy, which goes into
production of usual ``small'' NKG showers each having an initial
energy of $\Elpm$. Using this information, we can write the energy
loss equation:
\[ -\frac{d\En}{dt}=\frac{\En}{\sqrt{\En/\Elpm}}=\sqrt{\En\Elpm}
\qquad{\rm at\ }\En>\Elpm \]
where $t=z/X_0$ is the thickness in radiation lengths. Solving this
equation, we find the number of showers per unit length with starting
energy $\Elpm$:
\begin{eqnarray*}
\rho_{sm}(z)=-\frac{1}{\Elpm}\frac{d\En}{X_0dt}=
\frac{1}{X_0}\left(t_{\rm LPM}-\frac{t-t_i}{2}\right)\\
\textrm{for\ }t_i<t<t_i+2t_{\rm LPM}
\end{eqnarray*}
where $\tlpm=X_{\rm LPM, 0}/X_0=\sqrt{\En_0/\Elpm}$, and $t_i$
is the depth of the first interaction which can be taken $t_i=t_{\rm
LPM}$.  The number of particles in each ``small'' sub-shower can be
described approximately as
\[ N_{sm}(z)=N_{\rm max,LPM}e^{-\frac{(z-z_{\rm max})^2}{2L^2}} \]
where $L=1.5$~m as established in Section~\ref{sec:cherenkov} and $z_{\rm max}$
is the location of the maximum.
The maximum number of particles is given by~\cite[p.\ 23]{Sokolsky89}
\[ N_{\rm max}\approx3N_{e,\rm max}\approx
\frac{1}{\sqrt{\log(\En/\En_c)}}\frac{\En}{\En_c} \]
where $\En_c=(610{\rm\ MeV})/(Z+1.24)$ is the critical energy
\cite{passage}, which for water ($\langle Z\rangle=7.22$) is equal to
$72.1$~MeV.
At $\En=\Elpm$, we have $N_{\rm max,LPM}=\sci{8}{6}$. We assumed that
there are equal numbers of electrons, positrons and photons.
The total number of particle in the shower is given by a
convolution:
\[ N(z)=\int \rho_{sm}(z') N_{sm}(z-z')\,dz' \]
See comparison of this approximate theory and the results of Monte
Carlo calculations using program {\it LPMSHOWER}~\cite{Alvarez97} in
Figure~\ref{fig:lpmmodel}.

The charge excess \cite{Zas92} is estimated to be
\[\frac{N_{e^-}-N_{e^+}}{N_{e^-}+N_{e^+}}\approx 0.2\]
Thus $q(z)\approx 0.2e(2/3)N(z)$ where $e$ is the electron charge.
By the way, at $\En=1$~TeV, we have
$q_{\rm max}=\sci{9.6}{-17}$~C, almost the result of 
Section~\ref{sec:cherenkov} of $\sci{5.5}{-17}$~C.

Let us take its Fourier transform, $\tilde{N}(p)=\int
N(z)e^{-ipz}dz$. We use the fact that the Fourier transform of a
convolution is just the product of Fourier transforms.
The needed Fourier transforms are
\[ \tilde{\rho}_{sm}(p)=e^{-ipX_0(\tlpm+t_i)}\tlpm^2\left(
\frac{\sin\alpha}{\alpha}+\frac{i}{\alpha}\left[
\frac{\sin\alpha}{\alpha}-\cos\alpha
\right]\right)
\]
where $\alpha=pX_{\rm LPM, 0}$, and
\[\tilde{N}_{sm}(z)=e^{-ipz_{\rm max}}
N_{\rm max,LPM}\sqrt{2\pi}Le^{-(pL)^2/2} \]
For our purposes, the exact absolute phase is not important.
The electric field is
\[
R|\E(\omega)|=\mu\mu_0\tilde{q}\left(k[\cos\theta-1/n]\right)f\sin\theta
\]
At $p=0$ (i.e., Cherenkov angle), we
get the maximum value
\[ R|\E(\omega)|_{\rm max}\approx \sci{1.4}{-7}
\frac{\En_0}{1{\rm\ TeV}}\frac{f}{f_0}
{\rm\ V\,MHz^{-1}} \]
which is approximately the same as before. However, the width of
Cherenkov angle at high energies is determined by the extended length
of the shower. We can define it as the angle at which $|\E|$ is
reduced by a factor of $e^{-1/2}$, which occurs at
$\alpha=pX_{\rm LPM,0}\approx2$. We approximate
$p=k(\cos(\theta_c+\Delta\theta)-1/n)\approx (2\pi fn/c)
\sin\theta_c\,\Delta\theta$ and get the cone width due to LPM effect
to be
\[ \Delta\theta_{\rm LPM}\approx 0.9^\circ \frac{1}{\sqrt{\En_0/1{\rm\ EeV}}}
\frac{f_0}{f} \]
where $f_0=500$~MHz.

\section{Model-independent limit on differential flux}
\label{app:flux}
As we mentioned in Section~\ref{sec:sensitivity}, from a single
equation~(\ref{eq:intlim}) one cannot set in general a
model-independent limit on a differential flux $\Phi(\En)$. However,
after certain assumption of smoothness of function $\Phi(\En)$ it can
be done~\cite{Anchor02}.  Let us first consider a {\em
model-dependent} limit on differential flux $\Phi(\En)$ from a single
condition (\ref{eq:intlim}) assuming that $\Phi(\En)$ has a certain
functional form. Usually, it is assumed that $\Phi(\En)=\Phi_0
K(\En;P)$ where $K$ is a functional shape determined by a set of
parameters $P$. Then from (\ref{eq:intlim}) it follows that
\[ \Phi_0\le\frac{\smax}{\int\lambda(\En)K(\En;P)d\En} \]
or
\begin{equation}
\Phi(\En)\le\max_P \frac{\smax
  K(\En;P)}{\int\lambda(\En')K(\En';P)d\En'}
\label{eq:bestdifflim}
\end{equation}
It turns out that this equation is valid even when $\Phi(\En)$ is a
linear combination of functions $K(\En;P)$:
\[ \Phi(\En)=\int \Phi_0(P)K(\En;P)dP \]
We can prove it assuming the opposite.
If $\Phi(\En)\int\lambda(\En')K(\En';P)d\En'>\smax K(\En;P)$ for all $P$, then
\begin{eqnarray*}
\int \Phi(\En')\lambda(\En') d\En' =
\int \Phi_0(P) \left(\int \lambda(\En')K(\En';P)d\En'\right) dP\\
>\smax \frac{\int \Phi_0(P)K(\En;P) dP}{\Phi(\En)}=\smax
\end{eqnarray*}
which contradicts our initial assumption (\ref{eq:intlim}).

Although in this paper we will not use any concrete functions
$K(\En;P)$, we get a simple formula (\ref{eq:difflim}) from
(\ref{eq:bestdifflim}) by assuming that $K(\En;P)\equiv K(\En;\En_0)$
is a curve of width of $\agt \En_0$ centered at $\En_0$ and normalized
so that $\int K(\En;\En_0)d\En=1$. Then
$\max_{\En_0} K(\En;\En_0)$ is achieved at $\En_0\approx\En$ and is
$\alt 1/\En$, and if $\lambda(\En)$ is smooth enough, $\int
\lambda(\En')K(\En';\En_0)d\En\approx \lambda(\En_0)$, and is
$\approx\lambda(\En)$ when the expression on the right-hand side of
(\ref{eq:bestdifflim}) is maximized. So we estimate
\[ \Phi(\En)\alt \frac{\smax}{\En\lambda(\En)} \]

Thus, we have shown that a certain region of differential fluxes can
be rejected on the assumption that they are sufficiently smooth
functions of energy $\En$.

\section{Electric field at the satellite}
\label{app:efield}

\subsection{Transmission through ice}
\label{sapp:icetrans}
The Greenland ice
sheet at depths $<$1000~m has temperatures from $-25^\circ$~C to
$-20^\circ$~C \cite[pp. 23-24]{Bogorodsky85}.  The attenuation in ice
at frequency $f=35$~MHz at these temperatures is given
by~\cite{Johari75} and is $\approx$1~dB/100~m.

\subsection{Refraction}
\label{sapp:refraction}

First, let us find the refraction angle $r$. Consider satellite at
altitude $\hsat$, and the particle shower occurring at arc distance
$\ths$ from satellite position. Since the depth of the shower is
small compared to the satellite altitude, we can assume that
refraction also occurs at arc distance $\ths$.

The distance from the satellite to the refraction point is found using
cosine theorem:
\[ \Rsat=\sqrt{\Rearth^2+(\Rearth+\hsat)^2
-2\Rearth(\Rearth+\hsat)\cos\ths} \]
where $\Rearth$ is the Earth radius.  The nadir angle $\nadir$ is
found from $\sin\nadir=\Rearth\sin\ths/\Rsat$. The refraction angle
$r$ can be found from $\Rearth\sin r=(\Rearth+\hsat)\sin\nadir$, and
the incidence angle from Snell's law, $\sin i=\sin r/n$. Since
$i,r\in[0,\pi/2]$, $\cos i,r=\sqrt{1-\sin^2 i,r}$.

After the refraction at the ice surface, $\E$ changes according to the
Fresnel formulas \cite[pp.\ 281--282]{Jackson75}
\begin{eqnarray*}
\frac{E'_\perp}{E_\perp}&=&\frac{2n\cos i}{n\cos i+\cos r} \\
\frac{E'_\parallel}{E_\parallel}&=&\frac{2n\cos i}{n\cos r+\cos i}
\end{eqnarray*}
where $i$ and $r$ are the angles of incidence (from below) and
refraction, correspondingly, related by the Snell's law, $n\sin i=\cos
r$; $E$ and $E'$ are the incident (below the surface) and refracted
(above the surface) electric field components, $E_\perp$ and
$\E_\parallel$ are the components perpendicular and parallel to the
plane of incidence.

However, we also need to know how the waves diverge to be able to use
the expression for $R|\E|$. Let $R$ be the distance from the source to
the point at which refraction occurs. If we look from above the
surface, the waves diverge in such a way that they look like coming
from distance $R'$ below the surface. Then at the satellite, the field
is determined by relation $E_{\rm sat}\Rsat=E'R'$, where $\Rsat\gg R,R'$ is
the distance to the satellite. The inequality is well justified since
the shower occurs at depth $\sim$1~km in ice, while the satellite
altitude is 800~km. To find $R'$,
consider an area element $dA$ of the surface. Then
\[ dA=\frac{R^2d\Omega}{\cos i}=\frac{R'^2d\Omega'}{\cos r} \]
where $d\Omega$ is the solid angle element at which $dA$ is seen from
the source point and $d\Omega'$ gives divergence of rays emanating
from $dA$ above the surface. These solid angle elements are
$d\Omega=\sin i\,di\,d\phi$ and $d\Omega'=\sin
r\,dr\,d\phi$, where $\phi$ is azimuthal angle. Obtaining $di/dr$ from
Snell's law, we finally get
\[ R'=R\frac{\cos r}{n\cos i} \]
Thus, the modified Fresnel relations are
\begin{eqnarray*}
\frac{\Rsat E_{{\rm sat},\perp}}{RE_\perp}&=&\frac{2\cos r}{n\cos i+\cos r} \\
\frac{\Rsat E_{{\rm sat},\parallel}}{RE_\parallel}&=&
\frac{2\cos r}{n\cos r+\cos i}
\end{eqnarray*}

\subsection{Polarization and emission angles}
\label{sapp:polarization}

Although $E$ is given by (\ref{eq:zhs}), we need components $E_\perp$
and $E_\parallel$ to describe the refraction. Consider a particle
shower whose direction is described by a dip angle below horizon
$\dipang$ and azimuthal angle (in respect to the direction toward
satellite, calculated clockwise) $\phs$ (see
Figure~\ref{fig:view}). Introduce a coordinate system such that $z$
axis is vertical upward, $x$ axis is horizontal in the direction of
the satellite. Then the unit vector along the shower axis is
$\hat{a}=\{\cos\dipang\cos\phs,-\cos\dipang\sin\phs,-\sin\dipang\}$.
The unit vector in the direction of emission is $\hat{k}=\{\sin i, 0,
\cos i\}$.  The emission angle (between the shower axis and the
emission direction is found from $\cos\emang=\hat{a}\cdot\hat{k}$, so
that
\begin{eqnarray*}
\cos\emang&=&\cos\dipang\cos\phs\sin i-\sin\dipang\cos i\\
\sin\emang&=&\sqrt{1-\cos^2\emang}
\end{eqnarray*}

The polarization angle is the angle between the plane containing both
$\hat{a}$ and $\hat{k}$ and the $(x,z)$ plane. Consider
$\hat{h}=\hat{k}\times\hat{a}/|\hat{k}\times\hat{a}|$. Since
$|\hat{k}\times\hat{a}|=\sin\emang$,
{\setlength\arraycolsep{0pt}
\begin{eqnarray*}
\hat{h}=\frac{1}{\sin\emang}\{& &\cos i\cos\dipang\sin\phs,\\
& &\cos i\cos\dipang\cos\phs+\sin i\sin\dipang,\\
& &-\sin i\cos\dipang\sin\phs\}
\end{eqnarray*} }
The polarization angle is the angle between $\hat{h}$ and
$\hat{y}=\hat{z}\times\hat{x}$. Thus,
\[ \cos\polang=\hat{h}\cdot\hat{y}=
\frac{1}{\sin\emang}(\cos i\cos\dipang\cos\phs+\sin i\sin\dipang)
\]
and the sine is found from $\hat{h}\times\hat{y}=\hat{k}\sin\polang$, i.e.
\[
\sin\polang=\frac{\cos\dipang\sin\phs}{\sin\emang}
\]
Under this convention, the angle $\polang$ is calculated in the CCW
direction, when viewed from the source of the wave.

We can choose the polarization angle so that $\cos\polang>0$,
i.e. $\polang\in[-\pi/2,\pi/2]$, by adding to it $\pi$ when
$\cos\polang<0$. Then we get
\begin{eqnarray*}
\sin\polang&=&\frac{\cos\dipang\sin\phs}{\sin\emang}
\sign(\cos i\cos\emang+\sin\dipang) \\
\cos\polang&=&\sqrt{1-\sin^2\polang}
\end{eqnarray*}
where the argument of sign function has the same sign as the previous
expression for $\cos\polang$.

The electric field components are
\begin{eqnarray*}
E_\perp&=&E\sin\polang\\
E_\parallel&=&E\cos\polang
\end{eqnarray*}
so that the unit vectors in the directions of $E_\perp$, $E_\parallel$
and $k$ make a right-handed triad.

\subsection{Antenna response}
\label{sapp:antenna}

The analysis is based on information contained in \cite{Shao01}. FORTE
satellite has two antennas, A and B, perpendicular to each other and
the nadir direction. Antenna A is aligned with the ram (forward)
direction. Consider a signal coming from azimuthal direction $\azang$,
and at an angle $\nadir$ with nadir. Let us choose a coordinate system
so that $z$ axis is nadir, and the arrival direction is in $(x,z)$
plane. Then the antenna directions are given by
\begin{eqnarray*}
\hat{A}&=&\{\cos\azang,-\sin\azang,0\}\\
\hat{B}&=&\{\sin\azang,\cos\azang,0\}
\end{eqnarray*}
The signal arrival direction constitutes angles $\antang_A$ and
$\antang_B$ with the antennas, which are given by
\begin{eqnarray*}
\cos\antang_A=\hat{k}\hat{A}&=&-\sin\nadir\cos\azang\\
\cos\antang_B=\hat{k}\hat{A}&=&-\sin\nadir\sin\azang
\end{eqnarray*}

The electric field components parallel to antennas are
\begin{eqnarray*}
E_A&=&E_\parallel\cos\nadir\cos\azang-E_\perp\sin\azang\\
E_B&=&E_\parallel\cos\nadir\sin\azang+E_\perp\cos\azang
\end{eqnarray*}
We use values $E_A/\sin\antang_A$ and $E_B/\sin\antang_B$, which are
denoted as $E_x$ and $E_y$ in \cite{Shao01}, as inputs for the antenna
radiation diagrams (which are also found in~\cite{Shao01}) to get the
field recorded by the satellite.

\begin{acknowledgments}
This work was performed with support from the Los
Alamos National Laboratory's Laboratory Directed Research and
Development program, under the auspices of the United States
Department of Energy. PG is supported in part by a DOE OJI award
\#DE-FG 03-94ER40833.
\end{acknowledgments}

\bibliography{paper}

\end{document}